\documentclass[12pt]{article}
\usepackage{authblk}
\usepackage{url}

\usepackage{amsmath,amsfonts,amsthm}
\usepackage{mathtools}
\usepackage{array}

\usepackage{caption, ctable}
\usepackage[nameinlink]{cleveref}
\usepackage{colortbl}

\usepackage{amsmath,amsfonts,amsthm}
\usepackage{mathtools}
\usepackage{array}
\usepackage[caption=false,font=normalsize,labelfont=sf,textfont=sf]{subfig}
\usepackage{textcomp}
\usepackage{stfloats}
\usepackage{url}
\usepackage{graphicx}
\usepackage{cite}

\renewcommand{\o}{\text{\O}}
\newcommand{\st}{\,:\,}
\renewcommand{\mod}{\ \textup{mod}\,}

\newcommand{\mca}{$m$-dimensional Costas array}
\newcommand{\mpa}{$m$-dimensional permutation array}
\newcommand{\zm}{\mathbb{Z}^m}
\newcommand{\z}{\mathbb{Z}}

\DeclarePairedDelimiter\set{\{}{\}}
\DeclarePairedDelimiter\inprod{\langle}{\rangle}

\newtheorem{theorem}{Theorem}
\newtheorem{lemma}{Lemma}
\newtheorem{conjecture}{Conjecture}
\newtheorem{cor}{Corollary}
\newtheorem{prop}{Proposition}

\theoremstyle{definition}
\newtheorem{definition}{Definition}
\newtheorem{example}{Example}
\newtheorem{remark}{Remark}

\title{Multidimensional Costas Arrays and Their Periodicity}

\author[1]{Ivelisse Rubio}
\author[2]{Jaziel Torres}
\affil[1]{Department of Computer Science, University of Puerto Rico, R\'{\i}o Piedras}
\affil[2]{Department of Mathematics, University of Puerto Rico, R\'{\i}o Piedras}
\date{}                     
\begin{document}

\maketitle

\abstract{A novel higher-dimensional definition for Costas arrays is introduced.
This definition works for arbitrary dimensions and avoids some limitations of previous definitions.
Some non-existence results are presented for multidimensional Costas arrays preserving the Costas condition when the array is extended periodically throughout the whole space.
In particular, it is shown that three-dimensional arrays with this property must have the least possible order; extending an analogous two-dimensional result by H. Taylor.
Said result is conjectured to extend for Costas arrays of arbitrary dimensions.}




\section{Introduction}
Costas array is a permutation array, i.e., a square binary array with a single 1 per row and per column, with the property that the vectors joining pairs of 1's are all distinct, this being called the \emph{Costas condition} \cite{golomb-taylor1984} or \emph{Costas property} \cite{golomb1992t4}.
Costas arrays are useful in many applications, especially in radar/sonar detection and wireless communications \cite{golomb1982two,costas1984study,maric2006using}, and their study preserves contemporary validity as, to this day, their usefulness continues to find new applications \cite{correll2020costas,saikia2020costas,wang2019design,puranam2019amplitude}.
Costas arrays have also been an interesting object for mathematical research, with researchers looking at usual mathematical questions of existence, distribution, structure, constructions, and generalizations \cite{golomb2007status,drakakis2011open}.
For a comprehensive review on the history and basic theory of Costas arrays, see \cite{drakakis2006review}.
In this paper, we introduce a new multidimensional generalization of Costas arrays and study their periodicity, not only because it is an interesting mathematical inquiry, but because multidimensional analogs of Costas arrays are also useful in radar and optical communications \cite{ortiz2011three,munson2022radar}, digital watermarking \cite{moreno2011multi} and digital holography \cite{healy2015number}. 

\begin{figure}[htbp]
    \centering
    \includegraphics[width=80mm]{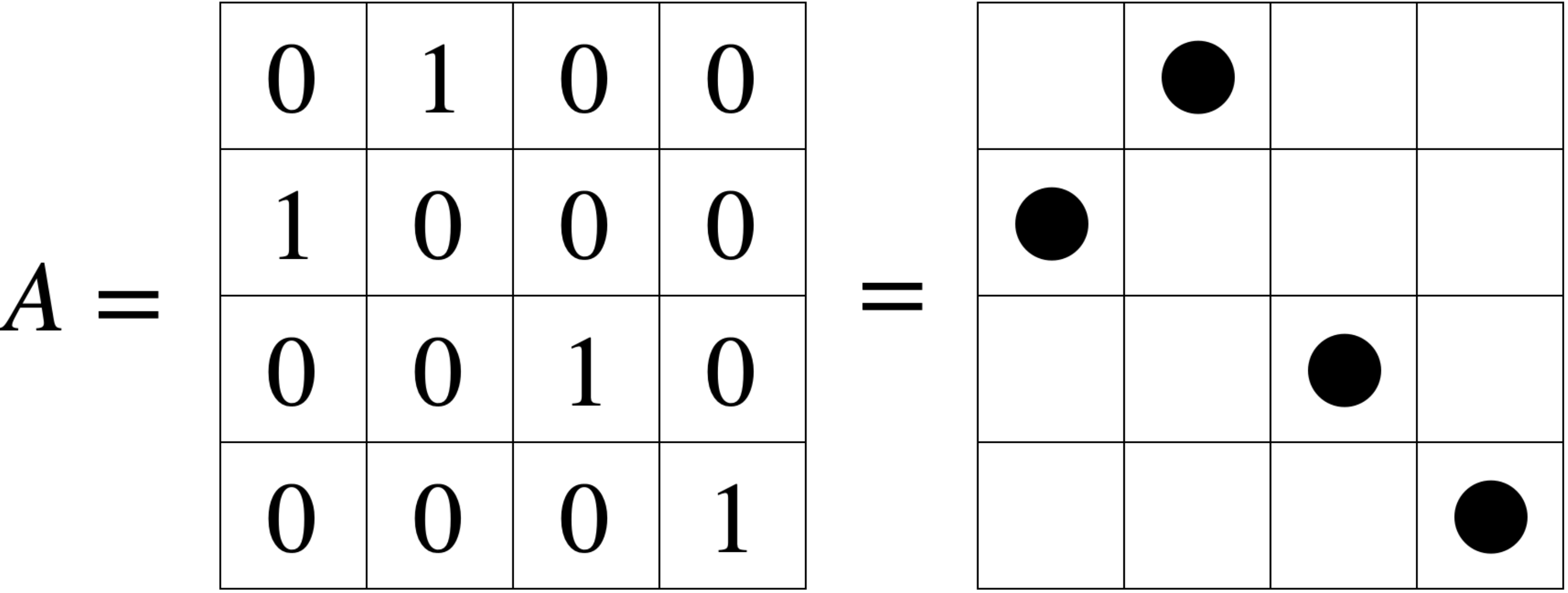}
    \caption{A Costas array of order 4}
    \label{fig:example1}
\end{figure}

 
 

To obtain a higher-dimensional analog of Costas arrays one has to generalize the two defining properties: being a permutation array and having no repeated difference vectors, i.e., the Costas condition.
Some multidimensional analogs of Costas arrays have been proposed before \cite{drakakis2008higher,healy2015number,jedwab2017costas,ortiz2013algebraic,batten2003permutations}, all satisfying the same multidimensional Costas condition, as it generalizes naturally; however, they differ in the generalization of a permutation array, as this can be done in different ways.
Nonetheless, the generalization in \cite[\S2]{healy2015number}, which produces arrays of the type defined in \cite[Definition 8]{drakakis2010generalization}, have an extremely low density of 1's, thus these arrays ``tend not to be very interesting'' \cite[p.~4]{drakakis2010generalization}.
The generalization in \cite[Definition 6]{drakakis2008higher}, further studied in \cite{drakakis2010generalization}, is problematic for odd dimensions.
The arrays in \cite[Definition 2]{jedwab2017costas} are only defined for three dimensions and their restriction to two dimensions do not produces a two-dimensional Costas array.
Finally, \cite[Definition 1]{ortiz2013algebraic} treats the arrays, and thus the vectors, over finite abelian groups, which is not consistent with the usual treatment of two-dimensional Costas arrays.
We propose a new multidimensional definition of Costas arrays that works for arbitrary dimensions, is consistent with the definition of a two-dimensional Costas array when restricted to two dimensions, and produces arrays with density of 1's equal to the square root of the number of entries. 
Moreover, \cite[Defnition 3.2]{batten2003permutations} and 
\cite[Definition 6]{drakakis2008higher} are special cases of our definition when restricted to permutations with one-dimensional domain and to arrays of even dimensions, respectively.

After introducing our definition, we study the existence of multidimensional arrays preserving the Costas condition when extended periodically to the whole space.
In this paper we focus on studying the higher-dimensional extensibility of the following result.
\begin{theorem}[{H. Taylor \cite{taylor1984non}}]\label{thm:taylor}
    For $n > 2$, let an $n \times n$ matrix of $n$ non-attacking rooks be extended doubly periodically over the whole plane.
    Then there must exist at least one $n\times n$ window in which some difference appears twice.
\end{theorem}
The \emph{non-attacking rooks} configuration in \Cref{thm:taylor} is equivalent to a permutation matrix.
It is clear that when any permutation array of order $n > 2$ is extended periodically to the whole plane, every $n\times n$ window contains a permutation array.
Nonetheless, \Cref{thm:taylor} is saying that in the periodic extension of a permutation array of order $n$ there is at least one $n \times n$ window that is not a Costas array, i.e., the Costas condition fails.
We show that an analogous result holds for three-dimensional arrays and for higher-dimensional arrays with odd number of 1's, and conjecture it holds for all higher-dimensional arrays.

Our motivation to study the existence of multidimensional arrays preserving the Costas condition and extending \Cref{thm:taylor} to higher-dimensional arrays is based on the early work in Costas arrays by S. W. Golomb, O. Moreno, and H. Taylor.
Firstly, Golomb and Taylor \cite{golomb-taylor1984}, by citing \Cref{thm:taylor}, stated that for $n > 2$, ``there does not exist a doubly periodic pattern with a Costas array in every $n\times n$ window'' (p.~1154), and pointed at the Welch construction as the closet to such configuration.
Then, Golomb and Moreno \cite{golomb1996periodicity} introduced \emph{circular Costas sequences}, which are equivalent to an $n \times n$ permutation matrix with the addition of an empty row, in which all the vectors joining pairs of 1's are distinct taken modulo the size of the array, i.e., modulo $n$ in their horizontal component and modulo $n+1$ in their vertical component.
The addition of the empty row was necessary because there are no Costas arrays with all vectors being distinct after taking modulo $n$ in both components.
Although a fairly simple pigeonhole argument works to see the latter, the existence of such array (with vectors distinct modulo $n$ in both components) would imply the existence of doubly periodic patters with a Costas array in every $n\times n$ window, which does not exist by \Cref{thm:taylor}.
Golomb and Moreno conjectured that the only circular Costas arrays are those from the Welch construction \cite[Conjecture 1]{golomb1996periodicity}, and this was proved by Muratovi\'c-Rubi\'c et al. in \cite[Theorem 3.4]{muratovic2015characterization}.
Our intention is to walk down and explore this chain of results on the periodicity of Costas arrays, but in the multidimensional context.
This paper is our first step as we explore a multidimensional analog of \Cref{thm:taylor}.

The rest of the paper is structured as follows.
In \Cref{sec:binary} preliminaries on multidimensional binary arrays are discussed, establishing all necessary definitions and notations.
In \Cref{sec:md_costas} a novel higher-dimensional definition of Costas arrays is introduced.
Lastly, \Cref{sec:periodicity} contains several non-existence results regarding the periodicity of Costas arrays.

\section{Preliminaries on Binary Arrays}\label{sec:binary}
Throughout the rest of this paper, $m$ is a natural number greater than 1. 

A \textbf{binary array} of dimension $m$ is a function $A:\Lambda\to\{0,1\}$ where $\Lambda$ is the hyper-rectangular subset of $\mathbb{N}^m$ given by
\[
    \Lambda = \{(a_1, a_2, \dots, a_m) \in \mathbb{N}^m \ : \ a_k \leq n_k, \ \ k=1,2, \dots, m \},
\]
for some natural numbers $n_1, n_2, \dots, n_m$.
Equivalently, $\Lambda = [n_1] \times [n_2] \times \cdots \times [n_m]$, where $[n]=\{1,2,\dots, n\}$.
We say that $\Lambda$ is the \textbf{index set} for the array $A$, and that $A$ has size $n_1 \times n_2 \times \cdots \times n_m$.
If in the index $\Lambda$, $n_i=1$ for some $i$, the $i$-th dimension of $A$ would be trivial, so we avoid those cases.
Hence, whenever we consider an $m$-dimensional array, we implicitly assume that the size in each dimension is at least 2.

For $\alpha = (a_1, \dots, a_m)\in\mathbb{Z}^m$, we denote by $\alpha_\Lambda$ the unique tuple $(a_1', \dots, a_m')\in\Lambda$ satisfying $a_i'\equiv a_i \pmod{n_i}$, $\forall i\in[m]$.
The \textbf{periodic extension} of $A$, denoted by $\mathbb{A}$, is the $m$-dimensional infinite array defined by
\[
    \mathbb{A}(\alpha) = A(\alpha_\Lambda), \qquad \forall \alpha \in \mathbb{Z}^m.
\]
In a binary array $A$, $\alpha\in\Lambda$ is called a \textbf{dot} of $A$ if $A(\alpha) = 1$.

For two distinct dots $\alpha = (a_1,\dots, a_m)$ and $\omega=(w_1,\dots,w_m)$ in an $m$-dimensional binary array $A$, the \textbf{difference vector} from $\alpha$ to $\omega$ is the vector

\[
    \omega-\alpha = \inprod{w_1-a_1, \dots, w_m-a_m} \in\zm.
\]
The \textbf{toroidal vector} \cite{jedwab2014deficiency} from $\alpha$ to $\omega$ is the vector
\[
   \inprod{(w_1 - a_1) \mod n_1, \ \dots, \ (w_m - a_m) \mod n_m}
     \in \z_{n_1}\times\cdots\times\z_{n_m}.
\]

To evade degenerate cases, whenever we consider difference or toroidal vectors, we assume the dots $\alpha$ and $\omega$ to be distinct, i.e., $\alpha\neq\omega$.
Notice our convention: we use parenthesis $(\cdot)$ for dots, and angled brackets $\inprod{\cdot}$ for vectors.

For each pair of dots in a binary array, there are two distinct difference vectors joining them: from $\alpha$ to $\omega$ and vice versa.
Hence, the number of difference vectors in a binary array with $n$ dots is $2 \binom{n}{2} = n(n-1)$, counting repetitions.
Similarly, the number of toroidal vectors is $n(n-1)$, counting repetitions.

\begin{definition}\label{def:multisets}
For an $m$-dimensional binary array $A$ of size $n_1 \times \cdots \times n_m$, define the following multisets (a set allowing repetitions):
\begin{itemize}
    \item $\mathcal{T}_A$ is the multiset of toroidal vectors occurring in $A$.
    \item $\mathcal{H}_A$ is the multiset of toroidal vectors $\inprod{h_1, \dots, h_m}$ occurring in $A$ for which $h_i = n_i/2$, for some $i \in [m]$.
\end{itemize}
\end{definition}

We use the letter ``H'' because a toroidal vector belongs to $\mathcal{H}_A$ if it has a component that is \emph{half} the length of the array in the corresponding direction.
As discussed before, for a binary array $A$ with $n$ dots, $|\mathcal{T}_A| = n(n-1)$.

\begin{lemma}\label{lemma:MDtoroidal_subarray}
    Let $A$ be an $m$-dimensional binary array with index set $\Lambda = [n_1] \times \cdots \times [n_m]$, and $\mathbb{A}$ its periodic extension to $\zm$.
    If $S$ is an $n_1 \times \cdots \times n_m$ window of $\mathbb{A}$, then $A$ and $S$ have the same multiset of toroidal vectors.
    That is, $\mathcal{T}_S = \mathcal{T}_A$.
\end{lemma}
\begin{proof}
    Let $S$ be an $n_1 \times \cdots \times n_m$ window of $\mathbb{A}$, and let $\psi$ be the function that maps every dot $\alpha \in S$ to the unique dot $\alpha_\Lambda \in A$.
    Since the window containing the array $S$ has the same size as the original array $A$, $\psi$ is a bijection from the dots of $S$ to the dots of $A$.
    By the definition of $\alpha_\Lambda$ and the definition of periodic extension,
    the toroidal vector from $\alpha$ to $\omega$ is equal to the toroidal vector from $\alpha_\Lambda$ to $\omega_\Lambda$.
    Hence, by the bijectiviy of $\psi$, $\mathcal{T}_S = \mathcal{T}_A$.
\end{proof}

\begin{prop}\label{prop:MDrepToroidal_implies_repDiff}
    Let $A$ be an $m$-dimensional binary array of size $n_1 \times \cdots \times n_m$ and $\mathbb{A}$ its periodic extension to $\zm$.
    If $A$ has a repeated toroidal vector $\inprod{h_1, \dots, h_m} \not\in \mathcal{H}_A$, then there is an $n_1 \times \cdots \times n_m$ window of $\mathbb{A}$ having a repeated difference vector.
\end{prop}
\begin{proof}
    Let $A$ be a binary array with index set $\Lambda = [n_1] \times \cdots \times [n_m]$, and let $\mathbb{A}$ be its periodic extension to $\mathbb{Z}^m$.
    Assume that $\langle h_1, \dots, h_m \rangle$ appears (at least) twice as a toroidal vector in $A$, with $0 \leq h_i \leq n_i-1$, for all $i\in [m]$.
    Then, there exist two pairs of dots of $A$, $\alpha_1 = (a_{11}, \dots, a_{1m}), \omega_1 = (w_{11},\dots, w_{1m})$ and $\alpha_2 = (a_{21},\dots, a_{2m}), \omega_2 = (w_{21}, \dots, w_{2m})$, such that
    \begin{align*}
        w_{1i}-a_{1i} \equiv w_{2i}-a_{2i} \equiv h_i \pmod{n_i}, \quad \forall i\in [m].
    \end{align*}
    
    We need to show that there are two pairs of dots of $\mathbb{A}$, 
    \[
        \alpha_1'=(a_{11}',\dots, a_{1m}'), \quad \omega_1'=(w_{11}',\dots, w_{1m}')\]
        \noindent and
    \[    \alpha_2'=(a_{21}',\dots, a_{2m}'), \quad \omega_2'=(w_{21}',\dots, w_{2m}')
    \]
    satisfying
    \begin{enumerate}
        \item[(i)] $w_{1i}'-a_{1i}' = w_{2i}'-a_{2i}'$, and
        \item[(ii)] $ k_i \leq a_{1i}', w_{1i}', a_{2i}', w_{2i}' \leq k_i + n_i-1, $ for some $k_i\in\mathbb{Z}$.
    \end{enumerate}
    Let us focus on the first coordinates.
    Notice that $a_{11}, w_{11} \in [n_1]$, hence $-(n_1-1) \leq w_{11}-a_{11} \leq n_1-1$.
    Therefore, $h_1 \equiv (w_{11}-a_{11}) \mod n_1$ implies $w_{11}-a_{11} = h_1$ or $w_{11}-a_{11} = h_1-n_1$.
    Similarly, $w_{21}-a_{21} = h_1$ or $w_{21}-a_{21} = h_1-n_1$.
    If $w_{11}-a_{11} = w_{21}-a_{21}$, choose $a_{11}' = a_{11}$, $w_{11}' = w_{11}$, $a_{21}' = a_{21}$, and $w_{21}' = w_{21}$.
    These four values satisfy (i) and (ii) with $k_1 = 0$.
    
    Otherwise, $w_{11}-a_{11} \neq w_{21}-a_{21}$ so one difference is equal to $h_1$ and the other one is equal to $h_1-n_1$, and also $h_1 > 0$.
    Without loss of generality, assume $w_{11}-a_{11} = h_1-n_1$ and $w_{21}-a_{21} = h_1$.
    Hence
    \begin{align}
        w_{11} - a_{11} < 0 &\implies w_{11} \leq a_{11}-1,\quad\textup{and}\label{ineq1} \\
        w_{21} - a_{21} > 0 &\implies a_{21} \leq w_{21}-1. \label{ineq2}
    \end{align}
    There are three cases: $a_{11} \leq w_{21}-1$, $w_{21} \leq a_{11}-1$ or $a_{11} = w_{21}$.
    
    \textit{Case 1.}
    If $a_{11} \leq w_{21}-1$, we set $a_{11}' = a_{11}$, $w_{11}' = w_{11}$, $a_{21}' = a_{21}$,  and $w_{21}' = w_{21}-n_1$.
    Notice that $w_{11}'-a_{11}' = w_{21}'-a_{21}' = h_1-n_1$, so (i) is satisfied.
    By the inequalities \eqref{ineq1} and \eqref{ineq2}, and the assumption $a_{11} \leq w_{21}-1$, (ii) is satisfied with $k_1 = w_{21}'$.
    
    \textit{Case 2.}
    If $w_{21} \leq a_{11}-1$, we set $a_{11}' = a_{11}-n_1$, $w_{11}' = w_{11}$, $a_{21}' = a_{21}$, and $w_{21}' = w_{21}$.
    In this case, $w_{11}'-a_{11}' = w_{21}'-a_{21}' = h_1$, so that (i) is satisfied.
    These values satisfy (ii) with $k_1 = a_{11}'$.
    
    \textit{Case 3.}
    If $w_{21} = a_{11}$ we have to consider two different cases:
    $h_1 < n_1/2$ or $n-h_1 < n_1/2$ (the case $h_1 = n_1/2$ does not happen by the hypothesis $\inprod{h_1, \dots, h_m} \not\in \mathcal{H}_A$).

    \textit{Case 3a.} If $h_1 < n_1/2$, we set $a_{11}' = a_{11}$, $w_{11}' = w_{11}+n_1$, $a_{21}' = a_{21}$, and $w_{21}' = w_{21}$.
    Condition (i) is satisfied because $w_{11}'-a_{11}' = w_{21}'-a_{21}' = h_1$.
    To see condition (ii), we note that $w_{21} = a_{11}$, and \eqref{ineq2} implies $a_{21} < w_{21} = a_{11} < w_{11}+n_1$.
    We have 
    \begin{align*}
        w_{11}'-a_{21}' &= (w_{11}' - a_{11}') + (a_{11}'-a_{21}')\\
        &= (w_{11}' - a_{11}') + (w_{21}'-a_{21}')  = h_1 + h_1.
    \end{align*}
    Therefore, $w_{11}'-a_{21}' \geq 0$, hence $a_{21}' \leq w_{11}'$.
    On the other hand, $w_{11}'-a_{21}' = h_1 + h_1 < n_1$ so that $w_{11}' < a_{21}' + n_1$.
    Using \eqref{ineq2} we conclude
    \[
        a_{21}' \leq a_{11}', w_{11}', a_{21}', w_{21}' \leq a_{21}'+n_1-1,
    \]
    and (ii) is satisfied with $k_1 = a_{21}'$.

    \textit{Case 3b.} If $n_1-h_1 < n_1/2$ we set $a_{11}' = a_{11}$, $w_{11}' = w_{11}$, $a_{21}' = a_{21}+n_1$, and $w_{21}' = w_{21}$.
    Condition (i) is satisfied because $w_{11}'-a_{11}' = w_{21}'-a_{21}' = h_1-n_1$.
    As in Case 3a, we have $w_{11} < a_{11} = w_{21} < a_{21} + n_1$.
    Also $a_{21}'-w_{11}' = (a_{21}'-w_{21}') + (a_{11}'- w_{11}') = (n_1-h_1) + (n_1-h_1) < n_1$, implying $w_{11}' \leq a_{21}'$ and $a_{21}' < w_{11}' + n_1$.
    Hence
    \[
        w_{11}' \leq a_{11}', w_{11}', a_{21}', w_{21}' \leq w_{11}'+n_1-1,
    \]
    and (ii) is satisfied with $k_1 = w_{11}'$.
    
    In a similar fashion we choose the remaining coordinates.
    At the end, by adding or subtracting $n_i$ to one of the $i$-th coordinates of the original points in $A$, namely $\alpha_1, \omega_1$ and $\alpha_2, \omega_2$, we obtained four dots $\alpha_1', \omega_1'$ and $\alpha_2', \omega_2'$ in $\mathbb{A}$ with equal difference vectors, and all four dots fitting in an $n_1 \times\cdots\times n_m$ window of $\mathbb{A}$.
\end{proof}

In the proof of \Cref{prop:MDrepToroidal_implies_repDiff}, the assumption that the repeated toroidal vector does not belong to $\mathcal{H}_A$ is only used in Case 3.
Hence, if a binary array has a repeated toroidal vector $\tau$ not falling under Case 3, we can follow the proof of \Cref{prop:MDrepToroidal_implies_repDiff} to obtain the same result, even when $\tau\in\mathcal{H}_A$.
We state this result as a corollary, which will be used in \Cref{thm:3d_costas_has_order_4} hereinbelow.

\begin{cor}\label{cor:md_Case3}
    Let $A$ be an $m$-dimensional binary array of size $n_1 \times \cdots \times n_m$ and $\mathbb{A}$ its periodic extension to $\zm$.
    Assume $A$ has two difference vectors $\omega_1-\alpha_1$ and $\omega_2-\alpha_2$ that are equal as toroidal vectors to $\inprod{h_1, \dots, h_m}$, where $\alpha_1=(a_{11}, \dots, a_{1m}), \ \omega_1=(w_{11}, \dots, w_{1m}), \ \alpha_2=(a_{21}, \dots, a_{2m}), \ \omega_2=(w_{21}, \dots, w_{2m})$ are dots of $A$.
    If for $i=1, \dots, m$,
    \begin{equation}\label{eq:md_condition_cor_case3}
    \begin{multlined}
        h_i=\frac{n_i}{2} \implies w_{1i}-a_{1i} = w_{2i}-a_{2i}, \quad\textup{or}\\ a_{1i}\neq w_{2i} \ \textup{and} \ a_{2i}\neq w_{1i},
    \end{multlined}
    \end{equation}
    then there is an $n_1 \times \cdots \times n_m$ window of $\mathbb{A}$ having a repeated difference vector.
\end{cor}

Notice that \Cref{prop:MDrepToroidal_implies_repDiff} has the flavor of a higher-dimensional analog of \Cref{thm:taylor}, but there is a subtle difference.
First and most obviously, \Cref{prop:MDrepToroidal_implies_repDiff} has the additional assumption that $\inprod{h_1, \dots, h_m} \not\in \mathcal{H}_A$. 
However, it is more general than \Cref{thm:taylor}, in the sense that it is stated for arbitrary binary arrays, not permutation arrays (the non-attacking rooks configuration).

\section{Multidimensional Costas Arrays}\label{sec:md_costas}

And now, the higher-dimensional generalization of Costas arrays we announced all along.
As discussed in the Introduction, the novelty of our definition resides in our definition of an \mpa.

\begin{definition}\label{def:md_perm}
    Let $\Lambda =  [n_1] \times \cdots \times [n_m]$.
    A binary array $A:\Lambda\to\set{0,1}$ is an \textbf{\mpa} if there is a bijection 
    \[
        \varphi: [n_1] \times \cdots \times [n_k] \to [n_{k+1}] \times \cdots \times [n_m],
    \]
    for some $k$, $1 \leq k < m$, such that, for $\alpha = (a_1, \dots, a_k, a_{k+1}, \dots, a_m) \in \Lambda$, $A(\alpha) = 1$ if and only if $\varphi(a_1, \dots, a_k) = (a_{k+1}, \dots, a_m)$.
    The \textbf{order} of a permutation array $A$, denoted $n$, is the number of dots in $A$, that is, $n = n_1n_2 \cdots n_k = n_{k+1}n_{k+2} \cdots n_m$.
\end{definition}

\begin{definition}\label{def:md_costas}
    An \textbf{$m$-dimensional Costas array} is an \mpa having no repeated difference vectors.
\end{definition}

\begin{figure}[ht]
    \centering
    \includegraphics[scale=.1]{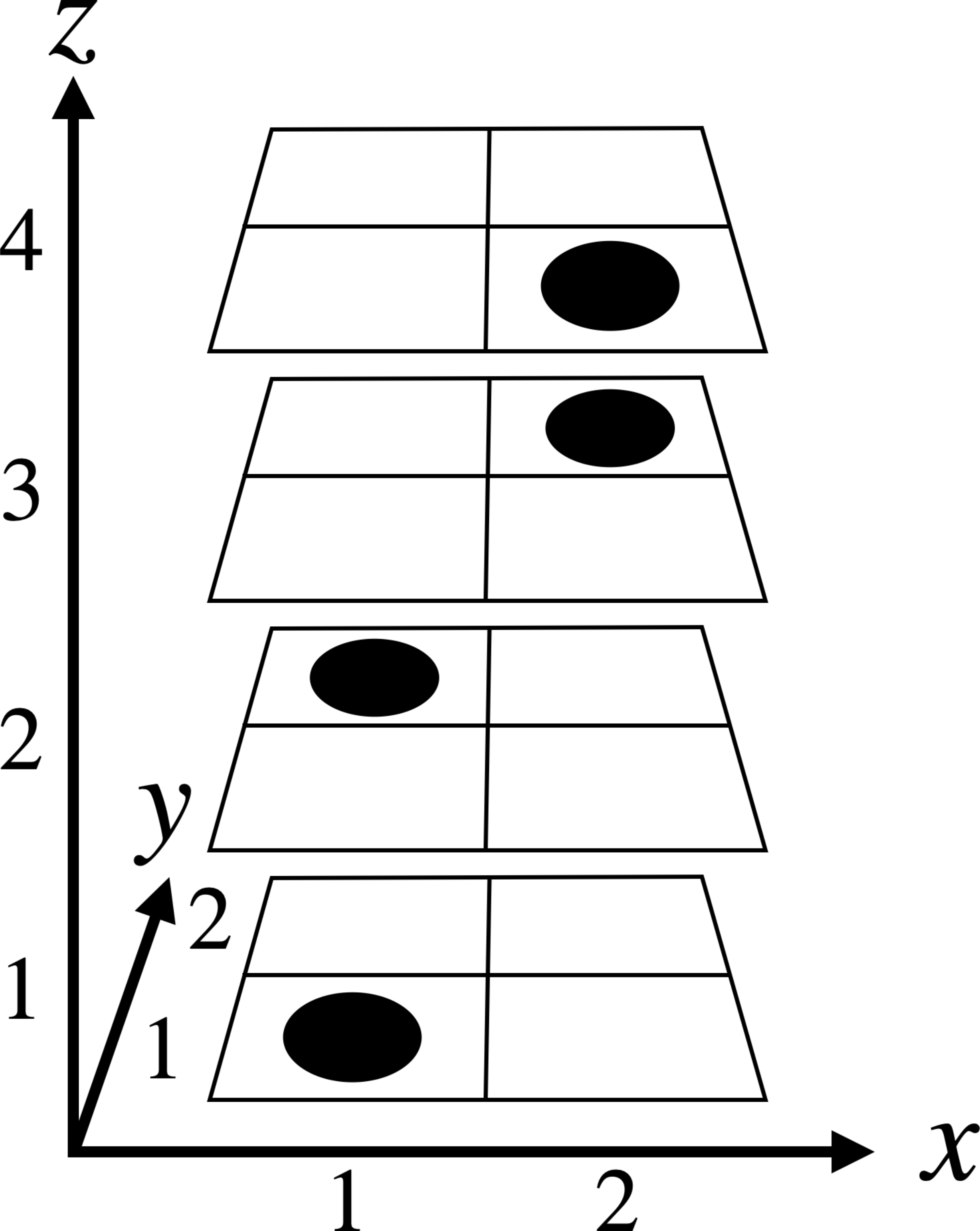}
    \caption{A three-dimensional Costas array of size $2 \times 2 \times 4$.}
    \label{fig:3d_costas}
\end{figure}

\begin{example}\label{ex1}
    Let $A:[2]\times[2]\times[4]\to\set{0,1}$ be the array with set of dots $\set{(1,1,1),$ $(1,2,2),$ $(2,2,3),$ $(2,1,4)}$.
    This array can be seen in \Cref{fig:3d_costas}.
    We can verify that this is a three-dimensional Costas array by computing all the difference vectors.
\end{example}

Notice that in \Cref{def:md_perm}, if $m=2$, then $k=1$, $\Lambda = [n_1]\times[n_2]$, and $\varphi$ is a bijection $\varphi:[n_1]\to[n_2]$, so $n_1 = n_2$.
Therefore, when $m=2$, the array $A$ in \Cref{def:md_costas} is a permutation array with no repeated difference vectors, which is exactly the definition of a two-dimensional Costas array.
If in \Cref{def:md_perm} we let $m$ to be even, $k=m/2$ and $n_1 = n_2 = \cdots = n_m$, a Costas array with this structure is precisely what is given in \cite[Defintion 6]{drakakis2008higher}.
Furthermore, if in \Cref{def:md_perm} we let $k=1$ so that $\varphi: [n_1]\to [n_2]\times\cdots\times[n_m]$ is a bijection, this special configuration is equivalent to a \emph{Costas array of dimension $m-1$ and type $n_2,\dots, n_m$}, as defined in \cite[Definition 3.2]{batten2003permutations}.
\Cref{def:md_costas} works for arbitrary dimensions, is consistent with the definition of two-dimensional Costas arrays when restricted to two dimensions, and produces arrays with square-root density: $n$ entries with 1's out of a total of $n^2$ entries.
The multidimensional analogs of Costas arrays proposed in \cite{drakakis2008higher,ortiz2013algebraic,healy2015number,jedwab2017costas} lack at least one of the aforementioned features.
A downside of our definition is that we do not know any systematic way of constructing multidimensional Costas arrays other than the \emph{reshaping} technique described in \cite[\S4]{drakakis2008higher} for the special case of arrays with $m$ even and $n_1 = n_2 = \cdots = n_m$.

To ease notation, from now on, let $X = [n_1] \times \cdots \times [n_k]$ and $Y = [n_{k+1}] \times \cdots \times [n_m]$, where the $n_i$'s are integers greater than 1, and $|X| = |Y|$. 

\begin{remark}\label{rmk1}
Having no repeated difference vectors in an \mca defined by a bijection $\varphi:X\to Y$ is equivalent to the so called \emph{distinct difference property}: for any $h\in\z^k$, $\varphi(i+h) - \varphi(i) = \varphi(j+h)-\varphi(j) \implies i = j$ or $h = (0,\dots, 0)$, for $i,i+h,j,j+h\in X$.
\end{remark}

If a bijection $\varphi: X \to Y$ defines an $m$-dimensional Costas array, the inverse map $\varphi^{-1}$ is also a bijection that defines an $m$-dimensional Costas array, since the dots of the latter would be just a swap between the first $k$ coordinates and the last $m-k$ coordinates of the former, so all difference vectors are going to be distinct. 
We consider those arrays to be equivalent. 
Moreover, if a bijection $\varphi: X \to Y$ defines an \mca, any permutation of the coordinates in $X$ or in $Y$ will produce another Costas array, as this only permutes the components of the difference vectors, so they are going to be distinct.
We consider those arrays to be equivalent.

\section{Periodicity of Multidimensional Costas Arrays}\label{sec:periodicity}
The non-existence of two-dimensional periodic patters preserving the Costas condition was settled by Taylor \cite{taylor1984non} with \Cref{thm:taylor}: there are no two-dimensional Costas arrays of order $n>2$, for which its periodic extension contains a Costas array in every $n\times n$ window.
Does the same happen for multidimensional Costas arrays?
Exploring this question is appropriate and relevant in the higher-dimensional context as there is no apparent reason why an analogous result should hold.

Without making any assessment of whether these arrays could exist, it is intuitive to consider the following two types of Costas periodicity, i.e., multidimensional arrays for which the Costas condition is preserved when periodically extending a multidimensional array.

\begin{definition}\label{def:md_periodic_costas}
    An \mca of size $n_1 \times \cdots \times n_m$ is \textbf{periodic Costas} if any $n_1 \times \cdots \times n_m$ window of its periodic extension has no repeated difference vectors, i.e, every window is an $m$-dimensional Costas array.
\end{definition}

\begin{remark}
    Based in \Cref{def:md_periodic_costas}, \Cref{thm:taylor} can be rephrased as: \textit{if $A$ is a two-dimensional periodic Costas array of order $n$, then $n=2$}.
\end{remark}

\begin{definition}\label{def:md_modular_costas}
    An \mca of size $n_1 \times \cdots \times n_m$ is \textbf{modular Costas} if any $n_1 \times \cdots \times n_m$ window of its periodic extension has no repeated toroidal vectors.
\end{definition}

\begin{remark}
    In \Cref{def:md_modular_costas} there is no need to consider the toroidal vectors in every window in the periodic extension because, by \Cref{lemma:MDtoroidal_subarray}, an \mca is modular Costas if and only if it has no repeated toroidal vectors.
\end{remark}

From the definitions follows that an \mca is periodic Costas if it is modular Costas.
However, these intuitive definitions are short lived, as we show in \Cref{thm:md_no_modular} that modular Costas does not exist, and conjecture an almost similar fate for periodic Costas, with \Cref{cor:odd_not_costas}, \Cref{thm:1d_image}, and \Cref{thm:3d_costas_has_order_4} supporting our conjecture (\Cref{conj:periodic} hereinbelow).
It is worth mentioning that, as with the addition of an empty row to define circular Costas arrays \cite{golomb1996periodicity}, one could modify multidimensional Costas arrays to define binary arrays, which are known to exists, that preserve the Costas condition periodically.
This is done by allowing an injection instead of a bijection in \Cref{def:md_perm}; see \cite[Chapter 3]{thesis} for further details and \cite{ortiz2013algebraic} for constructions.
However, our focus here is the multidimensional analog of \Cref{thm:taylor}.

To explore the periodicity of multidimensional Costas arrays, the sets defined next result to be quite useful.
\begin{definition}\label{def:sets}
Let $\Lambda = [n_1] \times \cdots \times [n_m]$ be an index set with $n_1 n_2 \cdots n_k$ $=$ $n_{k+1} n_{k+2}$ $\cdots$ $n_m$, for some $k<m$.
Define the following sets.
\begin{itemize}
	\item $\displaystyle T_\Lambda = (Z_1 \times Z_2)\setminus (Z_1\times\set{0} \cup \set{0}\times Z_2)$, where $Z_1 = \z_{n_1}\times \cdots \times \z_{n_k}$ and $Z_2 = \z_{n_{k+1}}\times \cdots \times \z_{n_m}$.
	\item $\displaystyle H_\Lambda = \set{\inprod{h_1, \dots, h_m}\in T_\Lambda \st h_i = n_i/2 \textup{ for some } i \in [m]}$.
\end{itemize}
\end{definition}

\begin{prop}\label{prop:value_set}
Let $A:\Lambda\to\set{0,1}$ be an \mpa. If $\tau$ is a toroidal vector in $A$, $\tau\in T_\Lambda$.
\end{prop}
\begin{proof}
    Let $\varphi:X \to Y$ be the bijection defining the dots of $A$, where $X = [n_1] \times \cdots \times [n_k]$ and $Y=[n_{k+1}] \times \cdots \times[n_m]$.
    Then $A$ is indexed by $\Lambda = X \times Y$.
    Let $Z_1 = \z_{n_1}\times \cdots \times \z_{n_k}$ and $Z_2 = \z_{n_{k+1}}\times \cdots \times \z_{n_m}$.
    If $\tau = \inprod{h_1, \dots, h_m}$ is a toroidal vector occurring in $A$, it is clear that $\tau\in Z_1 \times Z_2$.
    But, if $\tau$ is a toroidal vector from $\alpha = (a_1, \dots, a_m)\in A$ to $\omega =(w_1, \dots, w_m)\in A$ with $h_1 = h_2 = \cdots = h_k = 0$, it implies $a_1 = w_1$, $a_2 = w_2$, $\dots$, $a_{k} = w_k$.
    Therefore $\varphi(a_1, \dots, a_k) = \varphi(w_1, \dots, w_k) \implies (a_{k+1}, \dots, a_{m}) = (w_{k+1}, \dots, w_{m})$, so that $\alpha = \omega$. 
    In such case, $\tau$ is the zero vector, which we don't consider a valid toroidal vector.
    Similarly, for $\tau$ to be a valid toroidal vector in $A$, $h_{k+1} = \cdots = h_{m} = 0$ cannot happen.
    We conclude that $\tau\in T_\Lambda = (Z_1 \times Z_2)\setminus (Z_1\times\set{0} \cup \set{0}\times Z_2)$.
\end{proof}

As we can see from \Cref{prop:value_set}, the cardinality of the value set of toroidal vectors in a permutation array $A:\Lambda \to \set{0,1}$ of order $n$ is 
\begin{equation}\label{eq1}
    |T_\Lambda| = \left(\prod_{i=1}^k n_k - 1\right)\left(\prod_{i=k+1}^m n_k - 1\right) = (n-1)^2.
\end{equation}
Multidimensional permutation arrays of order $n$ have $n(n-1)$ toroidal vectors out of $(n-1)^2$ possible vectors.
Therefore, $\mathcal{T}_A$ has at least $n-1$ repeated elements, counting multiplicities.
We have proven the next result.

\begin{theorem}\label{thm:md_no_modular}
    Multidimensional modular Costas arrays do not exist.
\end{theorem}

\begin{figure}[ht]
    \centering
    \includegraphics[scale=.15]{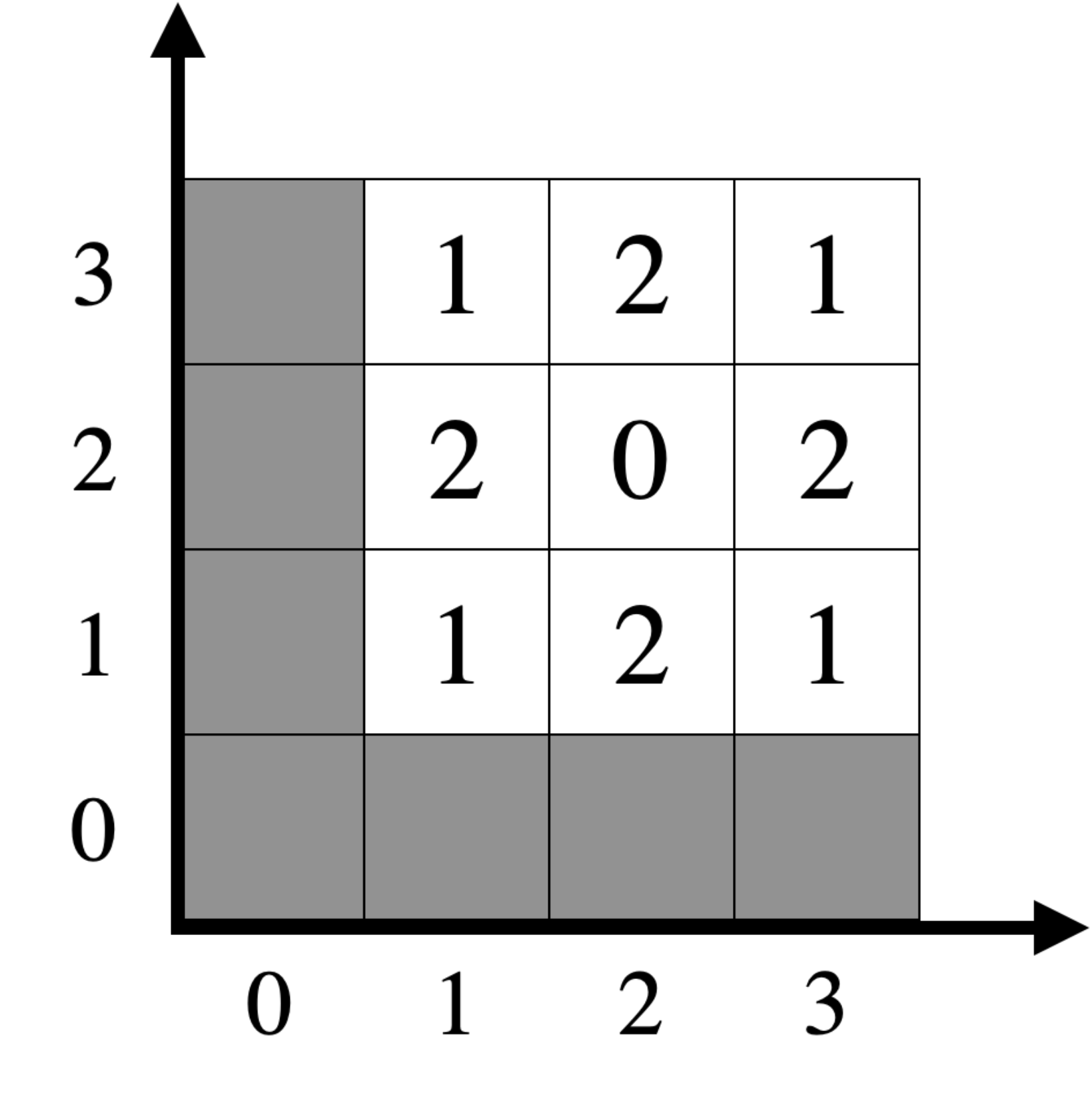}
    \caption{Frequency of toroidal vectors for the array in \Cref{fig:example1}}
    \label{fig:example2}
\end{figure}

\begin{figure}[ht]
    \centering
    \includegraphics[scale=.3]{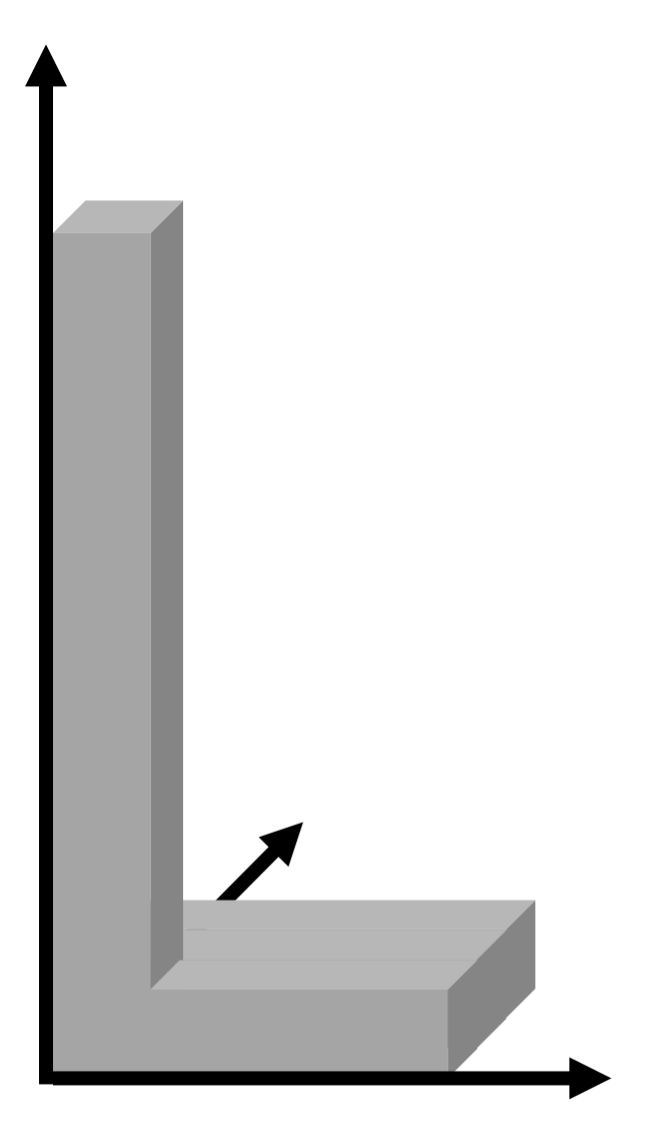}
    \caption{Shape of the excluded toroidal vectors in a three-dimensional Costas array.}
    \label{fig:example3}
\end{figure}

There is a nice picture to make sense of the multisets $\mathcal{T}_A$, $\mathcal{H}_A$ in \Cref{def:multisets} and the sets $T_\Lambda$, $H_\Lambda$ in \Cref{def:sets}.
Consider the permutation array in \Cref{fig:example1}, which is also Costas, and has toroidal vectors:
\begin{align*}
    \mathcal{T}_A = \{\{&\inprod{1,1}, \ \inprod{2,3}, \ \inprod{3,2}, \ \inprod{3,3}, \ \inprod{1,2}, \ \inprod{2,1},\\
    &\inprod{2,1}, \ \inprod{3,2}, \ \inprod{1,3}, \ \inprod{1,2}, \ \inprod{2,3}, \ \inprod{3,1}\}\},
\end{align*}
where we use double curly braces $\{\{ \cdot \}\}$ to denote a multiset.
Construct a frequency array (two-dimensional in this case), as in \Cref{fig:example2}, such that position $(x,y)$ contains the number of times $\inprod{x,y}$ appears as a toroidal vector in the Costas array in \Cref{fig:example1}.
Of course, an analogous frequency array can be constructed for any higher-dimensional permutation array.
The entries with coordinate $(0,\cdot)$ and $(\cdot, 0)$ are obscured because, in a permutation array, toroidal vector with those coordinates do not appear (those are the vectors in $Z_1\times\set{0} \cup\set{0}\times Z_2$; see \Cref{def:sets}).
In the case of a three-dimensional permutation array of size $n_1\times n_2\times n_1n_2$, the obscured entries (excluded toroidal vectors) of the frequency array are those of the form $(0,0,\cdot)$ and $(\cdot,\, \cdot, 0)$; something like the shaded region in \Cref{fig:example3}.

The set $T_\Lambda$ is the set of all toroidal vectors corresponding to the not excluded boxes (white boxes).
For our example in \Cref{fig:example1} and \Cref{fig:example2}, 

\[T_\Lambda = \{\inprod{1,1}, \inprod{1,2}, \inprod{1,3},\inprod{2,1},\inprod{2,2},\inprod{2,3},\inprod{3,1},\inprod{3,2},\inprod{3,3}\}.\]

\begin{figure}[ht]
    \centering
    \includegraphics[scale=.15]{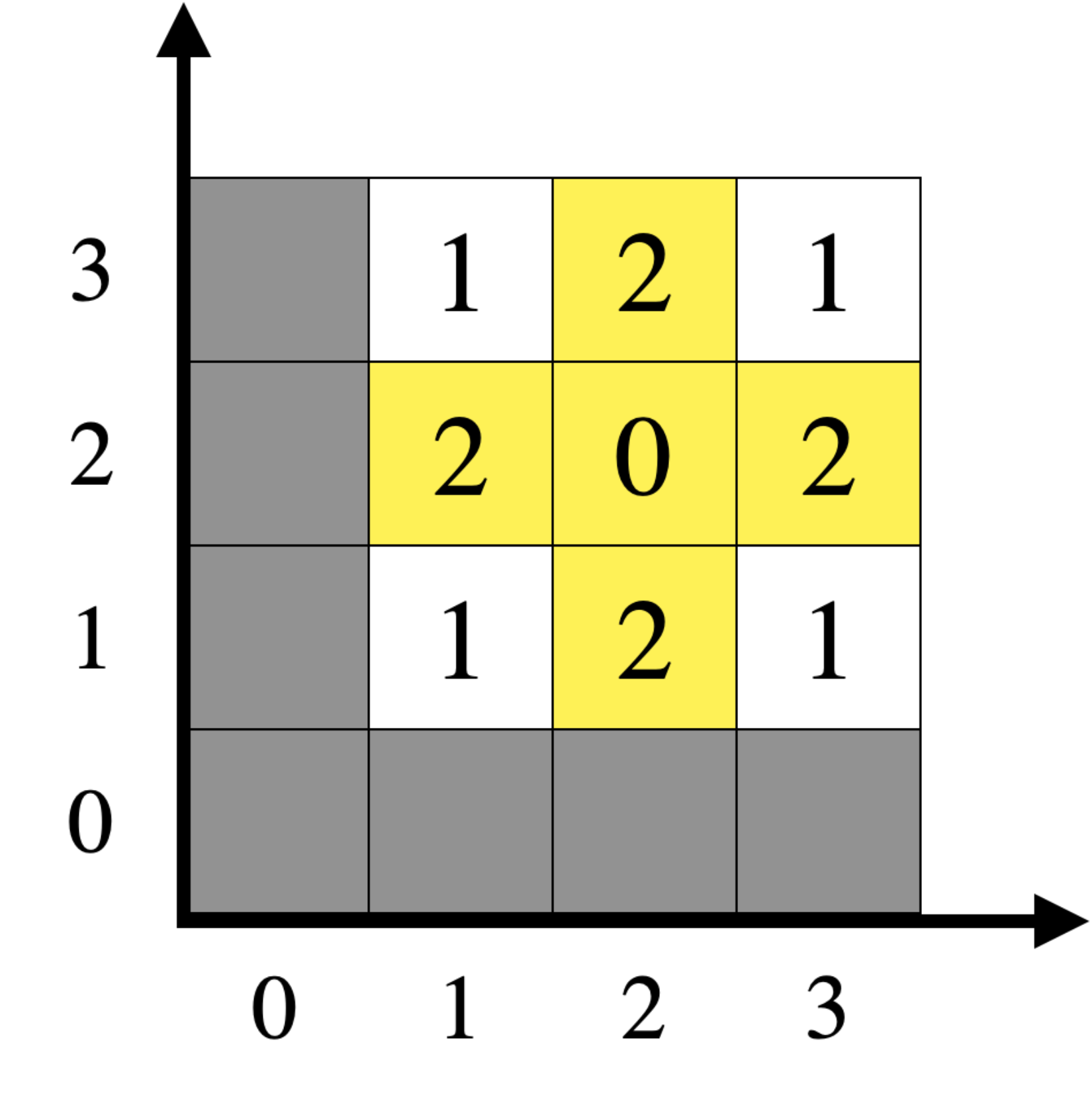}
    \caption{The frequency array in \Cref{fig:example3} with the entries corresponding to $H_\Lambda$ in yellow.}
    \label{fig:example4}
\end{figure}

The set $H_\Lambda$ represents the boxes in the frequency array corresponding to toroidal vectors with some component half the length of the matching side of the array.
We highlight those boxes in yellow.
For the array $A$ in \Cref{fig:example1}, whose order is $4$, those are the toroidal vectors with a component equal to 2, so we highlight column 2 and row 2, shown in \Cref{fig:example4}.
Therefore, $H_\Lambda = \set{ \inprod{1,2}, \ \inprod{2,1}, \ \inprod{2,2}, \ \inprod{2,3}, \ \inprod{3,2} }$.

By construction, $\mathcal{T}_A$ is the multiset of all toroidal vectors in the frequency array, each appearing repeatedly the number of times given by the number in the corresponding box. 
For example, the toroidal vector $\inprod{1,2}$ appears twice in $\mathcal{T}_A$, $\inprod{3,1}$ appears once, and $\inprod{2,2}$ does not appear in $\mathcal{T}_A$.
The multiset $\mathcal{H}_A$ contains the elements of $\mathcal{T}_A$ corresponding to yellow boxes in the frequency array.

Notice that any $4 \times 4$ permutation array will have the same boxes painted yellow as the ones in \Cref{fig:example4}.
That is the reason for our notation: we put $\Lambda$ as a subscript in $H_\Lambda$ to emphasize that the set only depends on the shape of the permutation array, not the entries, and the shape is determined by the index set $\Lambda$.
On the other hand, the numbers on the array with yellow boxes do depend on the permutation array $A$; thus, our subscript in $\mathcal{H}_A$.

After \Cref{thm:md_no_modular}, if there is any hope for the existence of arrays preserving the Costas condition periodically, it must rely solely on periodic Costas arrays, not modular Costas.
However, using the next lemma, we will show that multidimensional periodic Costas arrays do not exist for some classes of Costas arrays.

\begin{theorem}\label{big_lemma}
    Let $A$ be an \mpa of size $n_1 \times \cdots \times n_m$, index set $\Lambda = [n_1] \times \cdots \times [n_m]$, and order $n$.
    If $|\mathcal{H}_A| - |H_\Lambda|< n-1$, there is an $n_1 \times \cdots \times n_m$ window in the periodic extension of $A$ which has a repeated difference vector.
\end{theorem}
\begin{proof}
    We have
    \begin{align*}
        |\mathcal{H}_A| - |H_\Lambda| < n-1 \implies& |\mathcal{H}_A| - |H_\Lambda| < n(n-1) - (n-1)^2 \\
        \implies& (n-1)^2 - |H_\Lambda| < n(n-1) - |\mathcal{H}_A|\\
        \implies& |T_\Lambda-H_\Lambda| < |\mathcal{T}_A - \mathcal{H}_A|.
    \end{align*}
    The cardinality of the multiset $|\mathcal{T}_A - \mathcal{H}_A|$ is the number of toroidal vectors not in $\mathcal{H}_A$.
    Since the number of toroidal vectors of $A$ not in $\mathcal{H}_A$ is greater than the number of all possible values for toroidal vectors not in $H_\Lambda$, by the pigeonhole principle, $A$ must have a repeated toroidal vector not in $\mathcal{H}_A$.
    By \Cref{prop:MDrepToroidal_implies_repDiff}, an $n_1 \times \cdots \times n_m$ window of the periodic extension of $A$ has a repeated difference vector.
\end{proof}

Right away, we obtain a non-existence result for multidimensional periodic Costas arrays of odd order:
\begin{cor}\label{cor:odd_not_costas}
     If $A$ is an \mpa of size $n_1 \times \cdots \times n_m$ and odd order $n$, there is an $n_1 \times \cdots \times n_m$ window in the periodic extension of $A$ which has a repeated difference vector.
     In particular, multidimensional Costas arrays of odd order are not periodic Costas.
\end{cor}
\begin{proof}
    Notice that $n_i$ is odd for $i=1, \dots, m$ (otherwise the order of $A$ would be even), implying that $n_i/2$ is not an integer.
    Since the toroidal vectors in $A$ have integer components, none can have the $i$-th component equal to $n_i/2$.
    Then, $0 = |\mathcal{H}_A| - |H_\Lambda| < n-1$.
    By \Cref{big_lemma}, $A$ is not periodic Costas.
\end{proof}

Proving the non-existence of multidimensional periodic Costas arrays is more complicated for even order, at least with our approach.
The problem is that with an arbitrary permutation array of even order, we do not know how to obtain a sufficiently low upper bound for $|\mathcal{H}_A|$.
However, the following lemma will help us count some toroidal vectors in $\mathcal{H}_A$ for an \mpa $A$ of even order.

\begin{lemma}\label{lemma:counting}
    Let $A$ be an \mpa of even order defined by a bijection $\varphi: [n_1] \times \cdots \times [n_k] \to [n_{k+1}] \times \cdots \times [n_m]$, and $E \subseteq [m]$, where $n_i$ is even for all $i\in E$.
    Denote $\mathcal{H}_E$ the multiset of toroidal vectors $\inprod{h_1, \dots, h_m}$ occurring in $A$ for which $h_i = n_i/2$ for all $i \in E$.
    If $E \subseteq \set{1, \dots, k}$ or $E \subseteq \set{k+1, \dots, m}$, then
    \[
        |\mathcal{H}_E|\prod_{i\in E} n_i = n^2.
    \]
\end{lemma}
\begin{proof}
    Notice that $E \neq \o$ given that $A$ has even order.
    Assume $E \subseteq [k]$.
    After reordering indices we may assume $E = [t]$, for some $t \leq k$.
    For any $\alpha = (a_1, \dots, a_m) \in A$, set $w_i$ to be the unique number in $[n_i]$ that is equivalent to $a_i+n_i/2$ modulo $n_i$, for all $i\in[t]$.
    Choose $w_{t+1}, \dots, w_{k}$ freely and set $(w_{k+1}, \dots, w_m) = \varphi(w_1, \dots, w_k)$.
    By construction, $\omega = (w_1, \dots, w_m) \in A$, and if $\inprod{h_1, \dots, h_m}$ is the toroidal vector from $\alpha$ to $\omega$, $h_i = n_i/2$ for all $i\in E$.
    Hence, for each dot $\alpha\in A$, we can choose freely $w_{t+1}, \dots, w_{k}$ to obtain a toroidal vector in $A$ with $h_i = n_i/2$, for all $i\in E$.
    It is easy to see that this is the only way to obtain such toroidal vectors.
    By simple counting we have
    \begin{align*}
        |\mathcal{H}_E| &= n n_{t+1}\cdots n_k = n \frac{\prod_{i=1}^{k} n_i}{\prod_{i\in E} n_i} = n\frac{n}{\prod_{i\in E} n_i},
    \end{align*}
    and the result follows.
    The case $E \subseteq \set{k+1, \dots, m}$ is analogous.
\end{proof}

The next lemma provides a tidy formula for a counting, done by inclusion-exclusion, that will be used in \Cref{thm:1d_image} below.
It is proved by induction, and we omit the proof.

\begin{lemma}
    Let $K = \set{\set{n_1, \dots, n_k}}$ be a non-empty multiset of $k$ natural numbers.
    Then,
    \begin{equation}\label{eq:counting}
        \sum_{\substack{I \subseteq [k] \\ I\neq\o}} (-1)^{|I|+1} \frac{1}{\prod_{i\in I} n_i} = 1 - \prod_{i \in [k]}\frac{n_i-1}{n_i}.
    \end{equation}
\end{lemma}

We will tackle the existence of periodic Costas arrays of even order only for arrays defined by bijections with one-dimensional image, which are equivalent, by taking the inverse, to bijections with a one-dimensional domain.
We have two reasons for it.
First and foremost, the counting argument gets very convoluted for higher-dimensional image sets.
Secondly, every (non-equivalent) three-dimensional Costas array must have one-dimensional image, so the three-dimensional case will be covered.

\begin{theorem}\label{thm:1d_image}
    Let $A$ be an \mpa of even order $n$ defined by a bijection $\varphi: [n_1] \times \cdots \times [n_{m-1}] \to [n_m]$.
    Denote $\theta = 1 - \prod_{i\in E}\frac{n_i-1}{n_i}$, where $E = \set{i \in [m-1] \st n_i \textup{ is even}}$.
    If $\theta<\frac{n-2}{2n}$, there is an $n_1 \times \cdots \times n_m$ window in the periodic extension of $A$ which has a repeated difference vector.
    In particular, if $A$ is an \mca with $\theta<\frac{n-2}{2n}$, it is not periodic Costas.
\end{theorem}
\begin{proof}
    Let $n$ be the order of $A$.
    That is, $n = n_1n_2\cdots n_{m-1} = n_m$, which is even by assumption.
    By \Cref{big_lemma}, it is enough to check that $|\mathcal{H}_A| - |H_\Lambda| < n-1$.
    First we count $|\mathcal{H}_A|$.
    For it, define the following multisets:
    \begin{itemize}
        \item $\mathcal{U}$ is the multiset of toroidal vectors $\inprod{h_1, \dots, h_m}\in \mathcal{H}_A$ with $h_i = n_i/2$ for some $i \in [m-1]$.
        \item $\mathcal{V}$ is the multiset of toroidal vectors $\inprod{h_1, \dots, h_m}\in \mathcal{H}_A$ with $h_m = n_m/2$.
    \end{itemize}
    It is clear that $|\mathcal{H}_A| = |\mathcal{U}| + |\mathcal{V}| - |\mathcal{U} \cap \mathcal{V}|$, where the intersection is in the context of multisets, i.e, including repetitions.
    By the inclusion-exclusion principle $|\mathcal{U}|$ is the sum of the number of toroidal vectors with $h_i = n_i/2$ for a single $i\in[m-1]$, minus the toroidal vectors with $h_i = n_i/2$ for $i\in\set{i_1, i_2}\subseteq[m-1]$, plus the toroidal vectors with $i\in\set{i_1, i_2, i_3}\subseteq[m-1]$, and so on.
    By \Cref{lemma:counting}, for any $I \subset E$, the number of toroidal vectors with $h_i = n_i/2$, for all $i\in I$, is $\frac{n^2}{\prod_{i\in I} n_i}$. Therefore, 
    \[
    		|\mathcal{U}| =  \sum_{\substack{I \subseteq E \\ I\neq\o}} (-1)^{|I|+1} \frac{n^2}{\prod_{i\in I} n_i}.
    \]
    By \Cref{eq:counting}, $|\mathcal{U}| = n^2\theta$.
    Also by \Cref{lemma:counting}, 
    \[
        |\mathcal{V}| = \frac{n^2}{\prod_{i\in\set{m}} n_i} = n.
    \]
    Therefore, $|\mathcal{H}_A| \leq n(n\theta+1)$.
    
    To count $|H_\Lambda|$, define the following subsets of $H_\Lambda$:
    \begin{itemize}
        \item $\displaystyle U = \set{\inprod{h_1, \dots, h_m}\in H_\Lambda \st h_i = n_i/2 \textup{ for some } i \in [m-1]}$, and
        \item $\displaystyle V = \set{\inprod{h_1, \dots, h_m}\in H_\Lambda \st h_m = n_m/2}$.
    \end{itemize}
    Then $|H_\Lambda| = |U|+|V|-|U\cap V|$.
    Notice that, for fixed $i\in[m-1]$, there is a total of
    \[
        (n-1)\prod_{\substack{j \in [m-1] \\ j \neq i}} n_j = (n-1)\frac{n}{n_i}
    \]
    toroidal vectors $\inprod{h_1, \dots, h_m} \in H_\Lambda$ with $h_i = n_i/2$ (all entries are free to choose, except the $i$-th entry, which is fixed, and the $n-1$ factor is because $h_m$ could be any reminder modulo $n_m$, except zero).
    It follows from the inclusion-exclusion principle and \Cref{eq:counting} that
    \[
        |U| = (n-1)\sum_{\substack{I \subseteq E \\ I\neq\o}} (-1)^{|I|+1} \frac{n}{\prod_{i\in I} n_i} = (n-1)n\theta.
    \]
    On the other hand, $|V| = n-1$ because, for $\inprod{h_1, \dots, h_m} \in V$, $h_i, \dots, h_{m-1}$ can be chosen freely except for all zero.
    Finally, $|U\cap V| = \frac{|U|}{n-1} = n\theta$.
    Then,
    \[
        |H_\Lambda| = (n-1)n\theta + (n-1) - n\theta = (n-1)(n\theta+1)-n\theta.
    \]
    We conclude that
    \[
        |\mathcal{H}_A| - |H_\Lambda| \leq n(n\theta+1) - (n-1)(n\theta+1)+n\theta = 2n\theta+1.
    \]
    By assumption, $2n\theta+1 < n-1$, so the result follows from \Cref{big_lemma}.
\end{proof}

Notice that for a permutation array of even order defined by a bijection $\varphi: [n_1] \times \cdots \times [n_{m-1}] \to [n_m]$, if $n_i$ is very large for all $i\in[m]$, then $\theta$, as defined in \Cref{thm:1d_image}, is close to zero, while $\frac{n-2}{2n}$ is close to $1/2$, so we will have $\theta < \frac{n-2}{2n}$.
Therefore, sufficiently large Costas arrays defined by a bijection with one-dimensional image or one-dimensional domain are not periodic Costas.

The proof of \Cref{thm:1d_image} reveals another reason why we considered Costas arrays defined by a bijection with one-dimensional image.
Notice that the multisets $\mathcal{U}$ and $\mathcal{V}$ can be defined for arbitrary bijections, not only those having one-dimensional image.
That is, if $\varphi:[n_1] \times \cdots \times [n_k] \to [n_{k+1}]\times\cdots\times [n_m]$ defines a permutation array, define $\mathcal{U}$ as the multiset of toroidal vectors for which at least one of the first $k$ components is half the length of the array in the corresponding direction.
Similarly, define $\mathcal{V}$ as the multiset of toroidal vectors with at least one of the last $m-k$ components being half the length of the array in the corresponding direction.
As in the proof of \Cref{thm:1d_image}, $|\mathcal{H}_A| = |\mathcal{U}| + |\mathcal{V}| - |\mathcal{U} \cap \mathcal{V}|$.
The number $|H_\Lambda|$ is easy to count using the inclusion-exclusion principle.
Hence, to obtain $|\mathcal{H}_A|-|H_\Lambda|<n-1$, the hurdle is to get a sufficiently low upper bound for $|\mathcal{H}_A|$.
This can be done by obtaining a large lower bound on $|\mathcal{U} \cap \mathcal{V}|$, but this appears to be a difficult task.
Of course, $|\mathcal{U} \cap \mathcal{V}|\geq 0$ so zero is the worst possible lower bound.
When the image of $\varphi$ is one-dimensional, even the worst possible lower bound is good enough; zero is good enough.
When the image of $\varphi$ has a higher-dimensional image, zero is not in general a reasonable lower bound for $|\mathcal{U} \cap \mathcal{V}|$.

\begin{cor}\label{cor:3d_costas}
    Let $A$ be a three-dimensional permutation array of size $n_1\times n_2 \times n_1n_2$.
    Any of the following imply that there is an $n_1\times n_2 \times n_1n_2$ window in the periodic extension of $A$ having a repeated difference vector:
    \begin{enumerate}
        \item[(i)] $n_1$ and $n_2$ are odd.
        \item[(ii)] $n_1$ and $n_2$ are even, and one of them is greater than 4.
        \item[(iii)] $n_1$ is even greater than 2 and $n_2$ is odd, or vice versa.
    \end{enumerate}
\end{cor}
\begin{proof}
    Let $A$ be a permutation array defined by a bijection $\varphi:[n_1]\times[n_2]\to [n_1n_2]$; hence, the order of $A$ is $n = n_1n_2$.
    To avoid a degenerate three-dimensional array, we are implicitly assuming $n_1 > 1$ and  $n_2 > 1$.
    If (i) holds, $A$ has odd order and the result follows by \Cref{cor:odd_not_costas}. 
    
    Now assume $n_1$ and $n_2$ are even. 
    Then $\theta = 1 - \frac{(n_1-1)(n_2-1)}{n_1n_2}$ and
    \begin{align}
        \theta < \frac{n_1n_2-2}{2n_1n_2} 
            &\iff n_1n_2 - (n_1-1)(n_2-1) < \frac{n_1n_2-2}{2} \nonumber\\
            &\iff n_1 + n_2 - 1  < \frac{n_1n_2-2}{2} \nonumber\\
            &\iff 2 < \frac{n_1n_2}{n_1 + n_2}.\label{ineq3}
    \end{align}
    Inequality \eqref{ineq3} holds if $n_1>4$ or $n_2>4$.
    The result follows by \Cref{thm:1d_image}.
    
    Finally, assume $n_1$ is even, $n_2$ is odd, and $n_1 > 2$.
    In this case, $\theta = 1 - \frac{n_1-1}{n_1}$.
    Hence,
    \begin{align}
        \theta < \frac{n_1n_2-2}{2n_1n_2}
            &\iff n_1n_2 - n_2(n_1 - 1) < \frac{n_1n_2-2}{2} \nonumber\\
            &\iff 2n_2 < n_1n_2-2 \nonumber\\
            &\iff 2 < n_2(n_1-2). \label{ineq4}
    \end{align}
    But $n_2$ odd, $n_2>1$, and $n_1 > 2$ ensures that inequality \eqref{ineq4} holds.
    The result follows by \Cref{thm:1d_image}.
\end{proof}

With a bit more work, we can say even more than in \Cref{cor:3d_costas}.

\begin{theorem}\label{thm:3d_costas_has_order_4}
    If $A$ is a three-dimensional periodic Costas array, it has order 4.
\end{theorem}
\begin{proof}
    Let $A$ be a three-dimensional periodic Costas array.
    Without loss of generality, we assume $A$ is defined by a bijection $\varphi:[n_1]\times[n_2]\to[n_1n_2]$, where $n_1n_2$ is the order of $A$ and $n_1\leq n_2$.
    Since $A$ is periodic Costas, \Cref{cor:3d_costas} leaves only four possibilities:
    \begin{enumerate}
        \item[(i)] $n_1 = n_2 = 2$.
        \item[(ii)] $n_1 = 2$ and $n_2 = 4$.
        \item[(iii)] $n_1 = n_2 = 4$.
        \item[(iv)] $n_1 = 2$ and $n_2$ is odd.
    \end{enumerate}
    We must show (ii)--(iv) cannot happen.
    By exhaustive computation we checked that there are no periodic Costas arrays among all the $8!$ bijections $\varphi:[2]\times[4]\to[8]$ and all $16!$ bijections $\varphi:[4]\times[4]\to[16]$.
    
    Now we focus on case (iv).
    Assume $A$ is defined by a bijection $\varphi:[2]\times[k]\to[2k]$, for $k$ odd.
    Fix $(x_0,y_0)\in\z_{2}\times\z_k$, with $(x_0,y_0)\neq(0,0)$.
    Let $\alpha = (a_1,a_2,a_3)\in A$.
    If $\omega = (w_1,w_2,w_3)\in A$ is such that the toroidal vector from $\alpha$ to $\omega$ has the form $\inprod{x_0,y_0,z}$, for some $z\in\z_{2k}^*$, then
    \[
        w_1-a_1 \equiv x_0 \pmod{2} \qquad \textup{and} \qquad w_2-a_2 \equiv y_0 \pmod{n}.
    \]
    Since $w_1\in [2]$ and $w_2\in[k]$, their values are unique.
    But $A$ is defined by the bijection $\varphi$, so we must have $w_3 = \varphi(w_1,w_2)$.
    Therefore, for each $\alpha\in A$, we found a unique $\omega\in A$ such that the toroidal vector from $\alpha$ to $\omega$ has the from $\inprod{x_0,y_0,z}$, for some $z\in\z_{2k}^*$.
    We conclude that there are exactly $2k$ toroidal vectors $\omega$ of such form.
    However, there are only $2k-1$ possible choices for $z$ in  a toroidal vector of the form $\inprod{x_0,y_0,z}$.
    Thus, by the pigeonhole principle, for each pair $(x_0,y_0)\in(\z_2\times\z_k)^*$, there must be some $z_0 \in \z_{2k}^*$ such that $\inprod{x_0,y_0,z_0}$ is a repeated toroidal vector.
    In particular, let $x_0=0$, so there is a repeated toroidal vector with the form $\inprod{0,y_0,z_0}$.
    Notice that $y_0\neq k/2$ because $k$ is odd. 
    If $z_0\neq 2k/2 = k$, by \Cref{prop:MDrepToroidal_implies_repDiff}, $A$ is not periodic Costas, and the proof would be finished.
    
    Let $z_0 = k$.
    That is, assume $A$ has a repeated toroidal vector of the form $\inprod{0,y,k}$, for some $y\in\z_k^*$.
    We claim that this repeated toroidal vector will satisfy the conditions in \Cref{cor:md_Case3}, so $A$ is not periodic Costas.
    For the sake of a contradiction, assume $A$ has four dots
    \begin{align*}
         \alpha_1 &= (a_{11}, a_{12}, a_{13}),
        &\omega_1 &= (w_{11}, w_{12}, w_{13}),\\
         \alpha_2 &= (a_{21}, a_{22}, a_{23}),
        &\omega_2 &= (w_{21}, w_{22}, w_{23}),
    \end{align*}
    not satisfying the conditions in \Cref{cor:md_Case3} and for which $\omega_1-\alpha_1$ and $\omega_2-\alpha_2$ are equal as toroidal vectors to $\inprod{0,y,k}\in\z_2\times\z_k\times\z_{2k}$.
    Then, since $0\neq n_1/2 = 2/2 = 1$ and also $y \neq n_2/2 = k/2$ because $k$ is odd, if the four dots do not satisfy the conditions in \Cref{cor:md_Case3}, we must have
    $w_{13}-a_{13} = k$, $w_{23}-a_{23}=-k$, and $a_{13}=w_{23}$.
    Then, $w_{13}-a_{13} = -(w_{23}-a_{23}) = -(a_{13}-a_{23})$, implying $w_{13}=a_{23}$.
    $A$ is defined by the bijection $\varphi$, so all the dots of $A$ can be expressed as $(\varphi^{-1}(z),z)$, for some $z\in[2k]$.
    Hence, $\alpha_1 = (\varphi^{-1}(a_{13}), a_{13})$ and $\omega_2 = (\varphi^{-1}(w_{23}), w_{23})$, which implies $\alpha_1 = \omega_2$, given that $a_{13}=w_{23}$.
    Similarly, $\omega_1 = \alpha_2$ because $w_{13}=a_{23}$.
    Therefore, $\omega_1-\alpha_1$ and $\omega_2-\alpha_2 = \alpha_1-\omega_1$ are both equal as toroidal vectors to $\inprod{0,y,k}$.
  	
  	Given that $\omega_1-\alpha_1$ and $\omega_2-\alpha_2 = -(\omega_1 - \alpha_1)$ are both equal as toroidal vectors to $\inprod{0,y,k}$, $w_{12}-a_{12} \equiv y \pmod{k}$ and $-(w_{12}-a_{12}) \equiv y \pmod{k}$.
    Then, $2(w_{12}-a_{12}) \equiv 0 \pmod k$, but $k$ is odd and $a_{12},w_{12}\in[k]$, so $w_{12} = a_{12}$.
    Then we must have $y = 0$.
    This is a contradiction because $y\in\z_k^*$.
\end{proof}

The statement of \Cref{thm:3d_costas_has_order_4} raises a natural question.
Are there periodic Costas arrays of order $4$?
As the reader should expect, the answer is yes. 
Periodic Costas arrays of size $2\times 2\times 4$ do exist.
There are $4! = 24$ distinct bijections
\[
    \varphi:\set{(1,1),\ (1,2),\ (2,1),\ (2,2)} \to \set{1,2,3,4}.
\]
By exhaustive computation, we found that, out of these 24 bijections, 16 define Costas arrays, and 8 are periodic Costas.

\begin{example}
    Consider the three-dimensional Costas array described in \Cref{ex1} (see also \Cref{fig:3d_costas}).
    Although quite a task to do by hand, by checking all the 16 possibly distinct windows of size $2\times2\times4$ in its periodic extension, we can see that every window is a Costas array.
    Therefore, the array in \Cref{ex1} is a periodic Costas array.
\end{example}

Based on the above results and some exhaustive computations we performed, we finish this paper with a conjecture, which is a higher-dimensional analog of \Cref{thm:taylor}.

\begin{conjecture}\label{conj:periodic}
    Let $A$ be an $m$-dimensional Costas array of order $n$ defined by a bijection $\varphi:[n_1] \times \cdots \times [n_k] \longrightarrow [n_{k+1}]\times\cdots\times [n_m]$, where $k\geq m-k$.
    If $A$ is periodic Costas, $n = 2^k$.
    In particular, $n_1 = n_2 = \cdots = n_k = 2$.
\end{conjecture}
Notice that, by \Cref{thm:taylor} and \Cref{thm:3d_costas_has_order_4}, the above conjecture is true for two-dimensional and three-dimensional Costas arrays, respectively.

\section{Conclusion}
We proposed a new multidimensional generalization of Costas arrays.
Our definition works for arbitrary dimensions, the restriction to two-dimensions is consistent with the well known two-dimensional definition, produces arrays with density equal to the square root of the number of entries, and is more general than \cite[Defintion 6]{drakakis2008higher} and \cite[Definition 3.2]{batten2003permutations}.
We studied arrays whose periodic extension contains a Costas array in every window of the same size of the original array.
In the two-dimensional case, it was shown by H. Taylor \cite{taylor1984non} that those arrays must have order 2.
We showed partial results on the higher-dimensional extensibility of Taylor's theorem, and conjectured it holds for arbitrary dimensions.

With our definition for multidimensional Costas arrays there are as many research directions as there are for two-dimensional Costas arrays.
In fact, any result that is known for two-dimensional Costas arrays becomes a question in the higher-dimensional context.
We propose in \cite{thesis} a higher-dimensional analog to circular Costas arrays such that their relationship to multidimensional Costas arrays as defined in \Cref{def:md_costas} is consistent with the two-dimensional case.
The work includes results on the multidimensional extensibility of the Golomb-Moreno conjecture \cite{golomb1996periodicity}, shown in \cite{muratovic2015characterization} to be true for two-dimensional Costas arrays, thus validating \Cref{def:md_costas} as a feasible higher-dimensional definition of a Costas array.

A few interesting questions in the theoretical side for further directions:
Are there any systematic algebraic methods for constructing an \mca (cf. \cite{golomb1984algebraic})?
Does the proportion of Costas arrays among permutations decays exponentially as the size and/or the dimension increases (cf. \cite{warnke2021density})?
What can we say about the deficiency of multidimensional Costas arrays (cf. \cite{panario2011two,jedwab2014deficiency})? Are there any structural constrains for multidimensional Costas arrays (cf. \cite{jedwab2014structural,correll2018new})?

\section*{Acknowledgements}  This research was funded by the ``Fondo Institucional Para la Investigaci\'on (FIPI)'' from the University of Puerto Rico, R\'{\i}o Piedras.

\bibliographystyle{plain}
\bibliography{ref}

\begin{thebibliography}{10}

\bibitem{batten2003permutations}
Lynn~M. Batten and Sharad Sane.
\newblock Permutations with a distinct difference property.
\newblock {\em Discrete mathematics}, 261(1-3):59--67, 2003.

\bibitem{correll2018new}
Bill Correll.
\newblock A new structural property of {C}ostas arrays.
\newblock In {\em 2018 IEEE Radar Conference (RadarConf18)}, pages 0748--0753.
  IEEE, 2018.

\bibitem{correll2020costas}
Bill Correll, James~K. Beard, and Christopher~N. Swanson.
\newblock Costas array waveforms for closely spaced target detection.
\newblock {\em IEEE Transactions on Aerospace and Electronic Systems},
  56(2):1045--1076, 2020.

\bibitem{costas1984study}
John~P. Costas.
\newblock A study of a class of detection waveforms having nearly ideal
  range-{D}oppler ambiguity properties.
\newblock {\em Proceedings of the IEEE}, 72(8):996--1009, 1984.

\bibitem{drakakis2006review}
Konstantinos Drakakis.
\newblock A review of {C}ostas arrays.
\newblock {\em Journal of Applied Mathematics}, pages 4260--4265, 2006.

\bibitem{drakakis2008higher}
Konstantinos Drakakis.
\newblock Higher dimensional generalizations of the {C}ostas property.
\newblock In {\em 42nd Annual Conference on Information Sciences and Systems},
  pages 1240--1245. IEEE, 2008.

\bibitem{drakakis2010generalization}
Konstantinos Drakakis.
\newblock On the generalization of the {C}ostas property in higher dimensions.
\newblock {\em Advances in Mathematics of Communications}, 4(1):1--22, 2010.

\bibitem{drakakis2011open}
Konstantinos Drakakis.
\newblock Open problems in {C}ostas arrays.
\newblock arXiv Preprint arXiv:1102.5727, 2011.

\bibitem{golomb1984algebraic}
Solomon~W. Golomb.
\newblock Algebraic constructions for {C}ostas arrays.
\newblock {\em Journal of Combinatorial Theory, Series A}, 37(1):13--21, 1984.

\bibitem{golomb1992t4}
Solomon~W. Golomb.
\newblock The t4 and g4 constructions for {C}ostas arrays.
\newblock {\em IEEE Transactions on Information Theory}, 38(4):1404--1406,
  1992.

\bibitem{golomb2007status}
Solomon~W. Golomb and Guang Gong.
\newblock The status of {C}ostas arrays.
\newblock {\em IEEE Transactions on Information Theory}, 53(11):4260--4265,
  2007.

\bibitem{golomb1996periodicity}
Solomon~W. Golomb and Oscar Moreno.
\newblock On periodicity properties of {C}ostas arrays and a conjecture on
  permutation polynomials.
\newblock {\em IEEE Transactions on Information Theory}, 42(6):2252--2253,
  1996.

\bibitem{golomb1982two}
Solomon~W. Golomb and Herbert Taylor.
\newblock Two-dimensional synchronization patterns for minimum ambiguity.
\newblock {\em IEEE Transactions on Information Theory}, 28(4):600--604, 1982.

\bibitem{golomb-taylor1984}
Solomon~W. Golomb and Herbert Taylor.
\newblock Constructions and properties of {C}ostas arrays.
\newblock {\em Proceedings of the IEEE}, 72(9):1143--1163, 1984.

\bibitem{healy2015number}
John~J. Healy.
\newblock Number theoretical design of optimal holographic targets for
  measurement of 3d target motion.
\newblock In {\em 2015 International Conference on Optical Instruments and
  Technology: Optoelectronic Measurement Technology and Systems}, volume 9623,
  pages 257--262. SPIE, 2015.

\bibitem{jedwab2014deficiency}
Jonathan Jedwab and Jane Wodlinger.
\newblock The deficiency of {C}ostas arrays.
\newblock {\em IEEE Transactions on Information Theory}, 60(12):7947--7954,
  2014.

\bibitem{jedwab2014structural}
Jonathan Jedwab and Jane Wodlinger.
\newblock Structural properties of {C}ostas arrays.
\newblock {\em Advances in Mathematics of Communications}, 8(3):241--256, 2014.

\bibitem{jedwab2017costas}
Jonathan Jedwab and Lily Yen.
\newblock Costas cubes.
\newblock {\em IEEE Transactions on Information Theory}, 64(4):3144--3149,
  2017.

\bibitem{maric2006using}
Svetislav~V. Maric and Oscar Moreno.
\newblock Using {C}ostas arrays to construct frequency hop patterns for {OFDM}
  wireless systems.
\newblock In {\em 2006 40th Annual Conference on Information Sciences and
  Systems}, pages 505--507. IEEE, 2006.

\bibitem{moreno2011multi}
Oscar Moreno and Andrew~Z. Tirkel.
\newblock Multi-dimensional arrays for watermarking.
\newblock In {\em 2011 {IEEE} International Symposium on Information Theory
  Proceedings}, pages 2691--2695. IEEE, 2011.

\bibitem{munson2022radar}
Nicholas~R Munson, Travis~D Bufler, and Ram~M Narayanan.
\newblock Radar applications of orthogonal {S}udoku arrays and {C}ostas cubes.
\newblock In {\em Radar Sensor Technology XXVI}, volume 12108, pages 252--263.
  SPIE, 2022.

\bibitem{muratovic2015characterization}
Amela Muratovic-Ribic, Alexander Pott, David Thomson, and Qiang Wang.
\newblock On the characterization of a semi-multiplicative analogue of planar
  functions over finite fields.
\newblock {\em Topics in Finite Fields}, 632:317--326, 2015.

\bibitem{ortiz2011three}
Jos{\'e} Ortiz-Ubarri, Oscar Moreno, and Andrew Tirkel.
\newblock Three-dimensional periodic optical orthogonal code for {OCDMA}
  systems.
\newblock In {\em 2011 IEEE Information Theory Workshop}, pages 170--174. IEEE,
  2011.

\bibitem{ortiz2013algebraic}
Jos{\'e} Ortiz-Ubarri, Oscar Moreno, Andrew~Z. Tirkel, Rafael~A. Arce-Nazario,
  and Solomon~W. Golomb.
\newblock Algebraic symmetries of generic $(m+1)$-dimensional periodic {C}ostas
  arrays.
\newblock {\em IEEE Transactions on Information Theory}, 59(2):1076--1081,
  2013.

\bibitem{panario2011two}
Daniel Panario, Amin Sakzad, Brett Stevens, and Qiang Wang.
\newblock Two new measures for permutations: ambiguity and deficiency.
\newblock {\em IEEE Transactions on Information Theory}, 57(11):7648--7657,
  2011.

\bibitem{puranam2019amplitude}
Siva Sai~Krishna Puranam, Anand Gopinath, and Robert Sainati.
\newblock Amplitude based beam steering.
\newblock In {\em 2019 IEEE International Symposium on Phased Array System \&
  Technology (PAST)}, pages 1--4. IEEE, 2019.

\bibitem{saikia2020costas}
Monjul Saikia and Anwar Hussain.
\newblock Costas array based key pre-distribution scheme (cakps) for wsn and
  its performance analysis.
\newblock {\em CSI Transactions on ICT}, 8(3):347--354, 2020.

\bibitem{taylor1984non}
Herbert Taylor.
\newblock Non-attacking rooks with distinct differences.
\newblock Technical Report CSI-84-03-02, Communication Sciences Institute,
  University of Southern California, 1984.

\bibitem{thesis}
Jaziel Torres.
\newblock Multidimensional {C}ostas arrays and periodic properties.
\newblock Master's thesis, University of Puerto Rico, R\'io Piedras Campus,
  2022.

\bibitem{wang2019design}
Zhi Wang, Yanyong Su, Zhuofu Cui, and Xing Su.
\newblock A design of communication radar integrated signal of {MCPC} based on
  {C}ostas coding.
\newblock In {\em 2019 IEEE 4th International Conference on Image, Vision and
  Computing (ICIVC)}, pages 716--720. IEEE, 2019.

\bibitem{warnke2021density}
Lutz Warnke, Bill Correll~Jr., and Christopher~N. Swanson.
\newblock The density of {C}ostas arrays decays exponentially.
\newblock
  \url{https://mathweb.ucsd.edu/~lwarnke/CostasArrayExponentialDecay.pdf},
  2021.

\end{thebibliography}

\end{document}